\documentclass[twocolumn,showpacs,aps,prl,superscriptaddress]{revtex4}
\usepackage{graphicx}
\usepackage{dcolumn}
\usepackage{amsmath}
\usepackage{epsfig}

\RequirePackage{xspace}

\usepackage{relsize}
\def\babar{\mbox{\slshape B\kern-0.1em{\smaller A}\kern-0.1em
    B\kern-0.1em{\smaller A\kern-0.2em R}}}

\def\ccbar {\ensuremath{c\overline c}\xspace}
\def\piz   {\ensuremath{\pi^0}\xspace}

\def\pim   {\ensuremath{\pi^-}\xspace}

\def\Kbar  {\kern 0.2em\overline{\kern -0.2em K}{}\xspace}

\def\Kp    {\ensuremath{K^+}\xspace}

\def\Kpm   {\ensuremath{K^\pm}\xspace}

\def\KS    {\ensuremath{K^0_{\scriptscriptstyle S}}\xspace} 
\def\KL    {\ensuremath{K^0_{\scriptscriptstyle L}}\xspace} 
\def\Kstarz  {\ensuremath{K^{*0}}\xspace}

\def\Dstarp  {\ensuremath{D^{*+}}\xspace}
\def\B       {\ensuremath{B}\xspace}
\def\Bbar    {\kern 0.18em\overline{\kern -0.18em B}{}\xspace}

\def\Bz      {\ensuremath{B^0}\xspace}
\def\Bzb     {\ensuremath{\Bbar^0}\xspace}
\def\BzBzb   {\ensuremath{\Bz {\kern -0.16em \Bzb}}\xspace}
\def\Bu      {\ensuremath{B^+}\xspace}
\def\Bub     {\ensuremath{B^-}\xspace}

\def\Bpm     {\ensuremath{B^\pm}\xspace}

\def\BpBm    {\ensuremath{\Bu {\kern -0.16em \Bub}}\xspace}

\def\BorBbar    {\kern 0.18em\optbar{\kern -0.18em B}{}\xspace}
\def\DorDbar    {\kern 0.18em\optbar{\kern -0.18em D}{}\xspace}
\def\KorKbar    {\kern 0.18em\optbar{\kern -0.18em K}{}\xspace}

\def\jpsi     {\ensuremath{{J\mskip -3mu/\mskip -2mu\psi\mskip 2mu}}\xspace}
\def\psitwos  {\ensuremath{\psi{(2S)}}\xspace}

\def\etac     {\ensuremath{\eta_c}\xspace}

\def\chicone  {\ensuremath{\chi_{c1}}\xspace}

\mathchardef\Upsilon="7107
\def\Y#1S{\ensuremath{\Upsilon{(#1S)}}\xspace}

\def\FourS {\Y4S}

\mathchardef\Deltares="7101
\mathchardef\Xi="7104
\mathchardef\Lambda="7103
\mathchardef\Sigma="7106
\mathchardef\Omega="710A

\def\Deltabar{\kern 0.25em\overline{\kern -0.25em \Deltares}{}\xspace}
\def\Lbar{\kern 0.2em\overline{\kern -0.2em\Lambda\kern 0.05em}\kern-0.05em{}\xspace}
\def\Sigbar{\kern 0.2em\overline{\kern -0.2em \Sigma}{}\xspace}
\def\Xibar{\kern 0.2em\overline{\kern -0.2em \Xi}{}\xspace}
\def\Obar{\kern 0.2em\overline{\kern -0.2em \Omega}{}\xspace}
\def\Nbar{\kern 0.2em\overline{\kern -0.2em N}{}\xspace}
\def\Xb{\kern 0.2em\overline{\kern -0.2em X}{}\xspace}

\def\mes        {\mbox{$m_{\rm ES}$}\xspace}

\newcommand{\gev}{\ensuremath{\mathrm{\,Ge\kern -0.1em V}}\xspace}
\newcommand{\mev}{\ensuremath{\mathrm{\,Me\kern -0.1em V}}\xspace}
\newcommand{\gevc}{\ensuremath{{\mathrm{\,Ge\kern -0.1em V\!/}c}}\xspace}
\newcommand{\mevc}{\ensuremath{{\mathrm{\,Me\kern -0.1em V\!/}c}}\xspace}
\newcommand{\gevcc}{\ensuremath{{\mathrm{\,Ge\kern -0.1em V\!/}c^2}}\xspace}
\newcommand{\mevcc}{\ensuremath{{\mathrm{\,Me\kern -0.1em V\!/}c^2}}\xspace}

\def\ps         {\ensuremath{{\rm \,ps}}\xspace}  

\def\to                 {\ensuremath{\rightarrow}\xspace}
\newcommand{\stat}{\ensuremath{\mathrm{(stat)}}\xspace}
\newcommand{\syst}{\ensuremath{\mathrm{(syst)}}\xspace}
\def\pep2{PEP-II}
\def\BF{$B$ Factory}

\def\eps{\varepsilon\xspace}
\def\CP                {\ensuremath{C\!P}\xspace}

\def\stwob{\ensuremath{\sin\! 2 \beta   }\xspace}
\def\mistag{\ensuremath{w}\xspace}

\def\deltaz{\ensuremath{{\rm \Delta}z}\xspace}
\def\deltat{\ensuremath{{\rm \Delta}t}\xspace}
\def\deltamd{\ensuremath{{\rm \Delta}m_d}\xspace}

\newcommand{\jprBase}        {Phys.\ Rev.\xspace}

\newcommand{\nimBaseC}       {Nucl.\ Instr.\ and Methods\xspace}

\newcommand{\npBase}         {Nucl.\ Phys.\xspace}

\newcommand{\nim}       [1]  {\nimBaseC~{\bf #1}}

\newcommand{\np}        [1]  {\npBase\ {\bf #1}}
\newcommand{\pr}        [1]  {\jprBase\ {\bf #1}}

\newcommand{\progtp}    [1]  {{Prog.\ Theor.\ Phys.\ {\bf #1}}}

\def\leptontag{{\tt Lepton}}
\def\kaonitag{{\tt Kaon\,I}}
\def\kaoniitag{{\tt Kaon\,II}}
\def\kpitag{{\tt Kaon-Pion}}
\def\piontag{{\tt Pion}}
\def\othertag{{\tt Other}}
\def\syststwob{0.023}
\def\systal{0.013}

\newcommand{\bflav}{\ensuremath{\B_{{\rm flav}}}}

\newcommand{\BABARPubYear}    {04}
\newcommand{\BABARPubNumber}  {38}

\newcommand{\SLACPubNumber} {10652}

\def\figurebox#1#2#3{%
    \def\arg{#3}%
    \ifx\arg\empty
    {\hfill\vbox{\hsize#2\hrule\hbox to #2{\vrule\hfill\vbox to #1{\hsize#2\vfill}\vrule}\hrule}\hfill}%
    \else
    {\hfill\epsfbox{#3}\hfill}%
    \fi}

\begin{document}

\preprint{\babar-PUB-\BABARPubYear/\BABARPubNumber}
\preprint{SLAC-PUB-\SLACPubNumber}

\begin{flushleft}
\babar-PUB-\BABARPubYear/\BABARPubNumber\\
SLAC-PUB-\SLACPubNumber\\
\end{flushleft}

\title{
{\large \bf  Improved Measurement of 
{\boldmath\CP} Asymmetries in {\boldmath $\Bz \to (\ccbar) K^{0(*)}$} Decays}
}
%
\author{B.~Aubert}
\author{R.~Barate}
\author{D.~Boutigny}
\author{F.~Couderc}
\author{J.-M.~Gaillard}
\author{A.~Hicheur}
\author{Y.~Karyotakis}
\author{J.~P.~Lees}
\author{V.~Tisserand}
\author{A.~Zghiche}
\affiliation{Laboratoire de Physique des Particules, F-74941 Annecy-le-Vieux, France }
% ...
\collaboration{The \babar\ Collaboration}
\noaffiliation

\date{\today}


\begin{abstract}
We present results on time-dependent \CP
asymmetries in neutral $B$ decays to several \CP eigenstates. The
measurements use a data sample of about $227 \times 10^6$ $\FourS\to B\Bbar$
decays collected by the \babar\ detector at the
\pep2\ asymmetric-energy \BF\ at SLAC. 
The amplitude of the \CP asymmetry, \stwob\ in the Standard Model, 
is derived from decay-time distributions from events
in which one neutral $B$ meson is fully reconstructed in a final state
containing a charmonium meson and the other $B$ meson is determined 
to be either a \Bz or \Bzb from its decay products.
We measure $\stwob = 0.722  \pm 0.040 \stat \pm \syststwob \syst$ 
in agreement with the Standard Model expectation.
\end{abstract}


\maketitle
Charge-parity (\CP) violation in the $B$ meson system has been
established by the \babar~\cite{babar-stwob-prl} 
and Belle~\cite{belle-stwob-prl} collaborations.
The Standard Model of electroweak interactions describes \CP\ violation
as a consequence of an irreducible phase in the
three-generation Cabibbo-Kobayashi-Maskawa (CKM) quark-mixing
matrix~\cite{CKM}. In this framework, measurements of \CP\ asymmetries in 
the proper-time distribution of neutral $B$ decays to 
\CP\ eigenstates containing a charmonium and $K^{0}$ meson provide
a direct measurement of \stwob~\cite{BCP}. The angle 
$\beta$ is $\arg \left[\, -V_{\rm cd}^{}V_{\rm cb}^* / V_{\rm td}^{}V_{\rm tb}^*\, \right]$,
where $V_{ij}$ are CKM matrix elements.

In this Letter we report on an updated measurement of \stwob
in $(227 \pm 2) \times 10^6$ $B\Bbar$ decays using 
$\Bz$ decays to the final states
$\jpsi\KS$, $\jpsi\KL$, $\psitwos\KS$, $\chicone\KS$, $\eta_c \KS$, 
and $\jpsi\Kstarz (\Kstarz \to \KS\piz)$~\cite{chargeconj}.  
The \babar\ detector and the measurement technique 
are described in detail in Refs.~\cite{babar-detector-nim} and~\cite{babar-stwob-prd}, respectively.
Changes in the analysis with respect to the previously
published result include $140 \times 10^6$ more $B\Bbar$ events,
an improved event
reconstruction applied to all of the data,
a new flavor-tagging algorithm, and fewer
assumptions about the \CP\ properties of background events.

The proper-time distribution of $B$ meson decays 
to a \CP eigenstate $f$ 
can be expressed in terms of a complex parameter $\lambda$~\cite{lambda}, which
depends on both the \Bz-\Bzb oscillation amplitude and the 
decay amplitudes for $\Bzb \to f$ and $\Bz \to f$.
The decay rate ${\rm f}_+({\rm f}_-)$ when the 
other $B$ meson $B_{\rm tag}$ decays as a \Bz
(\Bzb) is given by
\begin{eqnarray}
{\rm f}_\pm(\, \deltat) = {\frac{e^{{- \left| \deltat \right|}/\tau_{\Bz} }}{4\tau_{\Bz} }}  
\Bigg[ 1 \Bigg.& \!\!\! \pm& \!\!\!  {\frac{2\mathop{\cal I\mkern -2.0mu\mit m}\lambda}{1 + |\lambda|^2} }
  \sin{( \Delta m_{d}  \deltat )} \nonumber \\
 &\!\!\! \mp& \!\!\! \Bigg. {\frac{1  - |\lambda|^2 } {1 + |\lambda|^2} }
       \cos{( \Delta m_{d}  \deltat) }  \Bigg],
\label{eq:timedist}
\end{eqnarray}
for a $B$ from a $\FourS \to \Bz \Bzb$ decay,
where $\Delta t$ is the difference between 
the proper decay times of the reconstructed $B$ meson $B_{\rm rec}$ and 
$B_{\rm tag}$,
$\tau_{\Bz}$ is the \Bz lifetime, and \deltamd is the
\Bz-\Bzb oscillation frequency.
The decay width difference $\Delta \Gamma$ between the \Bz\ mass eigenstates is assumed to be zero.
The sine term 
is due to the interference between direct
decay and decay after a net \Bz-\Bzb oscillation. 
A non-zero cosine term arises from the interference between 
decay amplitudes with different weak
and strong phases (direct \CP\ violation) or from \CP\ violation in 
\Bz-\Bzb mixing.
\par
In the Standard Model, \CP\ violation in mixing is negligible, as is direct \CP\
violation for 
$b \to \ccbar s$ decays that contain a charmonium meson~\cite{lambda}.  With these
assumptions $\lambda=\eta_f e^{-2i\beta}$,
where $\eta_f$ is the \CP
eigenvalue of final state $f$. Thus, the time-dependent
\CP asymmetry is
\begin{equation}
A_{\CP}(\deltat) \equiv  \frac{ {\rm f}_+  -  {\rm f}_- }
{ {\rm f}_+ + {\rm f}_- }
= -\eta_f \, \stwob \, \sin{ (\Delta m_{d} \, \deltat )} ,
\label{eq:asymmetry}
\end{equation}
with $\eta_f=-1$ for $\jpsi\KS$, $\psitwos\KS$, $\chicone \KS$, and $\eta_c \KS$, and
$+1$ for $\jpsi\KL$.
Due to the presence of even ($L$=0, 2) and odd ($L$=1) orbital angular momenta in the 
$B\to\jpsi\Kstarz$ final state, there can be \CP -even and \CP -odd contributions to the decay rate. 
When the angular information in the decay is ignored, the measured \CP\ asymmetry 
in $\jpsi\Kstarz$ is reduced by a factor $\vert 1-2R_{\perp} \vert$, where
$R_{\perp}$ is the fraction of the $L$=1 contribution. We have measured  
$R_{\perp} = 0.230 \pm 0.015 \pm 0.004$~\cite{BABARTRANS}, which 
gives an effective $\eta_f = 0.51 \pm 0.04$, after acceptance corrections. 

In addition to the \CP\ modes described above, 
we utilize a large sample (\bflav) of \Bz\ decays  to the flavor
eigenstates $D^{(*)-}h^+ (h^+=\pi^+,\rho^+$, and $a_1^+)$ and $\jpsi\Kstarz
(\Kstarz\to\Kp\pim)$ for calibrating our flavor
tagging and \deltat resolution.
Validation studies are performed with a control sample of $B^+$
mesons decaying to the final states $\jpsi K^{(*)+}$, $\psitwos
K^+$, $\chicone \Kp$,  and $\eta_c \Kp$.
The event selection and candidate reconstruction
are unchanged from those described in Refs.~\cite{babar-stwob-prl,babar-stwob-prd,etacks},
except that only the $\etac \to \KS \Kp \pim$ channel is used in the $\Bz \to \etac \KS$ and
$\Bpm \to \etac \Kpm$ modes ($2.91 < m_{\KS \Kp \pim} < 3.05 \gevcc$).

\par
The time interval \deltat between the two $B$ decays is calculated
from the measured separation \deltaz between the decay vertices of
$B_{\rm rec}$ and $B_{\rm tag}$ along the collision ($z$) axis~\cite{babar-stwob-prd}.
We find the $z$ position of the $B_{\rm rec}$ vertex from
its charged tracks. The $B_{\rm tag}$ decay vertex
is determined by fitting tracks not belonging to the $B_{\rm rec}$ 
candidate to a common vertex, employing constraints from the beam spot
location and the $B_{\rm rec}$ momentum~\cite{babar-stwob-prd}. 
We accept events with a calculated \deltat\ uncertainty of less than 2.5\ps
and $\vert \deltat \vert <20 \ps$.
The fraction of events satisfying these requirements is 95\%.
The r.m.s. \deltat resolution is 1.1\ps for the 99.7\% of these events that
exclude outliers.
\par
We use multivariate algorithms to identify signatures of $B$ decays that
determine (``tag'') the flavor at decay of the $B_{\rm tag}$ to be either a \Bz\ or \Bzb.
Primary leptons from semileptonic $B$ decays are selected
from identified electrons and muons
as well as isolated energetic tracks.
The charges of identified kaon candidates define a kaon tag.
Soft pions from \Dstarp decays are selected on the basis
of their momentum and direction with respect to
the thrust axis of $B_{\rm tag}$.
These algorithms are combined to
account for correlations among different sources of
flavor information and 
to provide an estimate of the mistag probability
for each event.
These algorithms have been improved relative to Ref.~\cite{babar-stwob-prl} 
with the addition of information
from low-momentum
electrons, $\Lambda \to p \pi$ decays, and additional correlations among
identified kaon candidates.
\par
Each event is assigned to 
one of six tagging categories 
if the estimated mistag probability is less than 45\%.
The \leptontag\ category contains events with an identified lepton;
the remaining events are divided into the 
\kaonitag, \kaoniitag, \kpitag, \piontag, or \othertag\ categories based on
the estimated mistag probability.  
This new 
definition of tagging categories improves
the overall performance
of the tagging algorithm, while largely preserving the 
separation of events with differing sources of tagging information.
For each category ($i$), the tagging efficiency $\eps_i$ and 
fraction $\mistag_i$ of events 
having the wrong tag assignment are measured from data 
(Table~\ref{tab:mistag}).
The effective tagging efficiency
$Q \equiv \sum_i {\eps_i (1-2\mistag_i)^2} $
improves 
by about 5\% (relative) over the algorithm used in
Ref.~\cite{babar-stwob-prl}. In addition, the correlations among
the mistag parameters and those of the \deltat\ resolution function are reduced.
\begin{table}[!t]
\caption
{Efficiencies $\epsilon_i$, average mistag fractions $\mistag_i$, mistag fraction differences
$\Delta\mistag_i \equiv \mistag_i(\Bz)-\mistag_i(\Bzb)$, and $Q$ extracted for each tagging
category $i$ from the $B_{\rm flav}$ sample. 
}
\label{tab:mistag} 
\begin{ruledtabular} 
\begin{tabular*}{\hsize}{l
@{\extracolsep{0ptplus1fil}}  D{,}{\ \pm\ }{-1} 
@{\extracolsep{0ptplus1fil}} D{,}{\ \pm\ }{-1} 
@{\extracolsep{0ptplus1fil}}  D{,}{\ \pm\ }{-1} 
@{\extracolsep{0ptplus1fil}}  D{,}{\ \pm\ }{-1}}  
Category     & 
\multicolumn{1}{c}{$\ \ \ \varepsilon$   (\%)} & 
\multicolumn{1}{c}{$\ \ \ \mistag$       (\%)} & 
\multicolumn{1}{c}{$\ \ \ \Delta\mistag$ (\%)} &
\multicolumn{1}{c}{$\ \ \ Q$             (\%)} \\ \colrule  
  \leptontag &   8.6,0.1&   3.2,0.4 &  -0.2,0.8&    7.5,0.2 \\
   \kaonitag &  10.9,0.1&   4.6,0.5 &  -0.7,0.9&    9.0,0.2 \\
  \kaoniitag &  17.1,0.1&  15.6,0.5 &  -0.7,0.8&    8.1,0.2 \\
     \kpitag &  13.7,0.1&  23.7,0.6 &  -0.4,1.0&    3.8,0.2 \\
    \piontag &  14.5,0.1&  33.0,0.6 &   5.1,1.0&    1.7,0.1 \\
   \othertag &  10.0,0.1&  41.1,0.8 &   2.4,1.2&    0.3,0.1 \\
\colrule
         All &  74.9,0.2&           &          &   30.5,0.4 \\
\end{tabular*} 
\end{ruledtabular} 
\end{table} 
\begin{figure}[!b]
\begin{center}%
\epsfig{figure=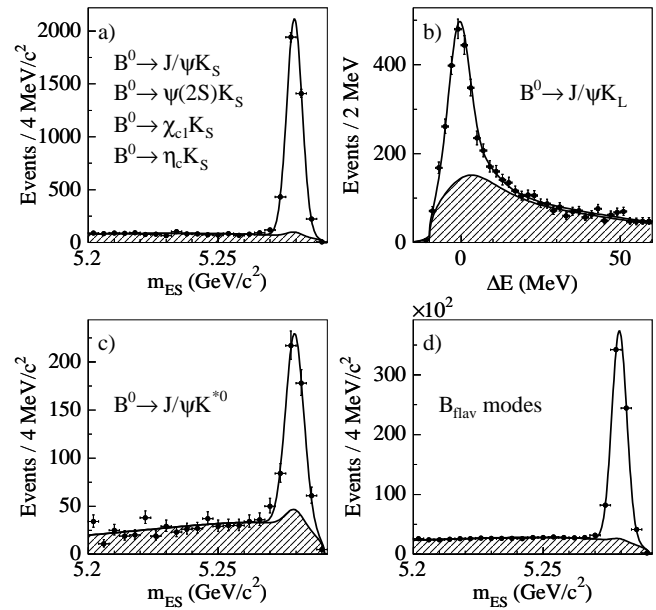, width=\linewidth}
\caption{
Distributions for $B_{\CP}$ and \bflav\ candidates satisfying the tagging and vertexing requirements:
a) \mes\ for the final states $J/\psi\KS $, $\psi(2S)\KS$, $\chi_{c1}\KS$,
and $\etac\KS$, 
b) $\Delta E$ for the final state $\jpsi\KL$,
c) \mes\ for $J/\psi K^{*0}(K^{*0}\to \KS\pi^0)$, and d) 
\mes\ for the \bflav\ sample. In each plot, the shaded region is the estimated background contribution.}
\label{fig:bcpsample}
\end{center}
\end{figure}
\begin{table}[!htb] 
\caption{ 
Number of events $N_{\rm tag}$ in the signal region after tagging and vertexing requirements, 
signal purity $P$ including the contribution from peaking background,
and results of fitting for \CP\ asymmetries in the $B_{\CP}$ sample and various subsamples.
In addition, results on the $B_{\rm flav}$ and charged $B$ control samples
test that no artificial \CP\ asymetry is found where 
we expect no \CP\ violation ($\stwob=0$).
Errors are statistical only. The 
signal region is $5.27 < \mes < 5.29 \gevcc$ ($\vert \Delta E \vert < 10\mev$ for $\jpsi \KL$).
}
\label{tab:result} 
\begin{ruledtabular} 
\begin{tabular*}{\hsize}{ l@{\extracolsep{0ptplus1fil}} r c@{\extracolsep{0ptplus1fil}} D{,}{\ \pm\ }{-1} } 
 Sample  & $N_{\rm tag}$ & $P(\%)$ & \multicolumn{1}{c}{$\ \ \ \stwob$}
\\ \colrule
 Full \CP\ sample                                   & $7730$        & $76$       &  0.722,0.040   \\ 
\hline
$\jpsi\KS$,$\psitwos\KS$,$\chicone\KS$,$\etac\KS$   & $4370$        & $90$       &  0.75, 0.04   \\
$\jpsi \KL$                                         & $2788$        & $56$       &  0.57, 0.09   \\
$\jpsi\Kstarz (\Kstarz \to \KS\piz)$                 & $572$         & $68$       &  0.96, 0.32   \\
\hline
1999-2002 data                                       &  $3032$        &  $77$         &  0.74 ,0.06   \\
2003-2004 data                                       &  $4698$        &  $77$         &  0.71 ,0.05   \\
\hline
\hline
\multicolumn{4}{l}{$\jpsi\KS$, $\psitwos\KS$, $\chicone\KS$, $\etac\KS$ only  $(\eta_f=-1)$ }  \\
\hline
$\ \ \jpsi \KS$ ($\KS \to \pi^+ \pi^-$)    & $2751$        & $96$       &  0.79, 0.05 \\
$\ \ \jpsi \KS$ ($\KS \to \pi^0 \pi^0$)    & $653$        & $88$       &  0.65, 0.12 \\
$\ \ \psi(2S) \KS$ ($\KS \to \pi^+ \pi^-$) & $485$        & $82$       &  0.88, 0.14 \\
$\ \ \chicone \KS $                        & $194$         & $81$       &  0.69, 0.23 \\
$\ \ \etac\KS $                            & $287$        & $64$       &  0.17, 0.25 \\
\hline
$\ $ \leptontag\ category                & $490$        & $96$       &  0.75, 0.08   \\
$\ $ \kaonitag\ category                 & $648$        & $93$       &  0.75, 0.08   \\
$\ $ \kaoniitag\ category                & $1021$        & $89$       &  0.77, 0.09   \\
$\ $ \kpitag\ category                   & $769$        & $90$       &  0.77, 0.15   \\
$\ $ \piontag\ category                  & $835$        & $87$       &  0.96, 0.22   \\
$\ $ \othertag\ category                 & $607$        & $88$       &  0.23, 0.51   \\
\hline\hline
$B_{\rm flav}$ sample                    & $72878$      & $85$       &  0.021, 0.013   \\
\hline 
$B^+$ sample                             & $18294$      & $88$       &  0.003, 0.020   \\
\end{tabular*} 
\end{ruledtabular} 
\end{table}

The beam-energy substituted mass
$\mes=\sqrt{{(E^{\rm cm}_{\rm beam})^2}-(p_B^{\rm cm})^2}$ 
(all modes except for $\jpsi\KL$)
or the difference $\Delta E$ between the candidate center-of-mass energy and
$E^{\rm cm}_{\rm beam}$ ($\jpsi \KL$ channel) 
are used to determine the composition of our final sample
(Fig.~\ref{fig:bcpsample}).
Here, $E^{\rm cm}_{\rm beam}$ and  $p_B^{\rm cm}$ are the 
beam energy and $B$ momentum in the center-of-mass frame.
Events with $\mes > 5.2 \gevcc$ ($\Delta E < 80\mev$) are
used so that the properties of the background contributions
can be measured.
The more restricted signal region (Table~\ref{tab:result}) 
contains 7730 \CP\ candidate events that satisfy the tagging and 
vertexing requirements.

For all modes except $\eta_c \KS$ 
and $\jpsi\KL$ we use simulated events to estimate the fractions of
events that peak in the $\mes$ signal region
due to cross-feed from other decay modes (peaking background).
For the $\eta_c\KS$ mode the cross-feed fraction is determined from a fit
to the $M_{KK\pi}$ and \mes\ distributions in data.
For the $\jpsi\KL$ decay mode, the composition, 
effective 
$\eta_f$, 
and
$\Delta E$ distribution of the individual background sources are
determined either from simulation (for $B\to\jpsi X$) 
or from the $m_{\ell^+ \ell^-}$ sidebands in data (for fake $\jpsi\to \ell^+ \ell^-$).
\par
We determine \stwob with a simultaneous maximum likelihood fit
to the \deltat distributions of the tagged $B_{\CP}$ and \bflav\
samples. The \deltat\ distributions of the $B_{\CP}$ sample are modeled by
Eq.~\ref{eq:timedist} with $|\lambda|=1$.
Those of the \bflav\ sample evolve
according to the known frequency for flavor oscillation in $B^0$
mesons. The observed amplitudes for the \CP asymmetry in the
$B_{\CP}$ sample and for flavor oscillation in the \bflav\ sample
are assumed to be reduced by the same factor $1-2\mistag$ due to flavor mistags.
The \deltat distributions for the signal are
convolved with a common resolution function, modeled by
the sum of three Gaussians~\cite{babar-stwob-prd}.
Backgrounds are incorporated with an empirical
description of their \deltat spectra, containing prompt and 
non-prompt components convolved with a resolution
function~\cite{babar-stwob-prd} distinct from that of the signal.
\par
There are 65 free parameters in the fit: \stwob (1),
the average mistag fractions $\mistag$ and the
differences $\Delta\mistag$ between \Bz\ and \Bzb\ mistag fractions for each
tagging category (12), parameters for the signal \deltat resolution (7),
parameters for \CP\ background time dependence (8), and the difference between
$\Bz$ and $\Bzb$ reconstruction and tagging efficiencies (7); for 
\bflav\ background, time dependence (3), \deltat resolution
(3), and mistag fractions (24). For the \CP\ modes (except for $\jpsi \KL$), 
the apparent \CP\ asymmetry of the non-peaking background in each tagging
category is allowed to float.  This asymmetry is parameterized
so that it does not depend on the value of \stwob.

We fix $\tau_{\Bz}=1.536\ps$, $\deltamd
=0.502\ps^{-1}$~\cite{PDG2004},
$\vert \lambda \vert = 1$, and $\Delta \Gamma=0$.
The determination of the mistag fractions and \deltat resolution
function parameters for the signal is dominated by the high-statistics \bflav\ sample. 
Background parameters are determined mainly from events with
$\mes < 5.27\gevcc$. 
\par
\begin{figure}[ht]
\begin{center}
 \epsfig{figure=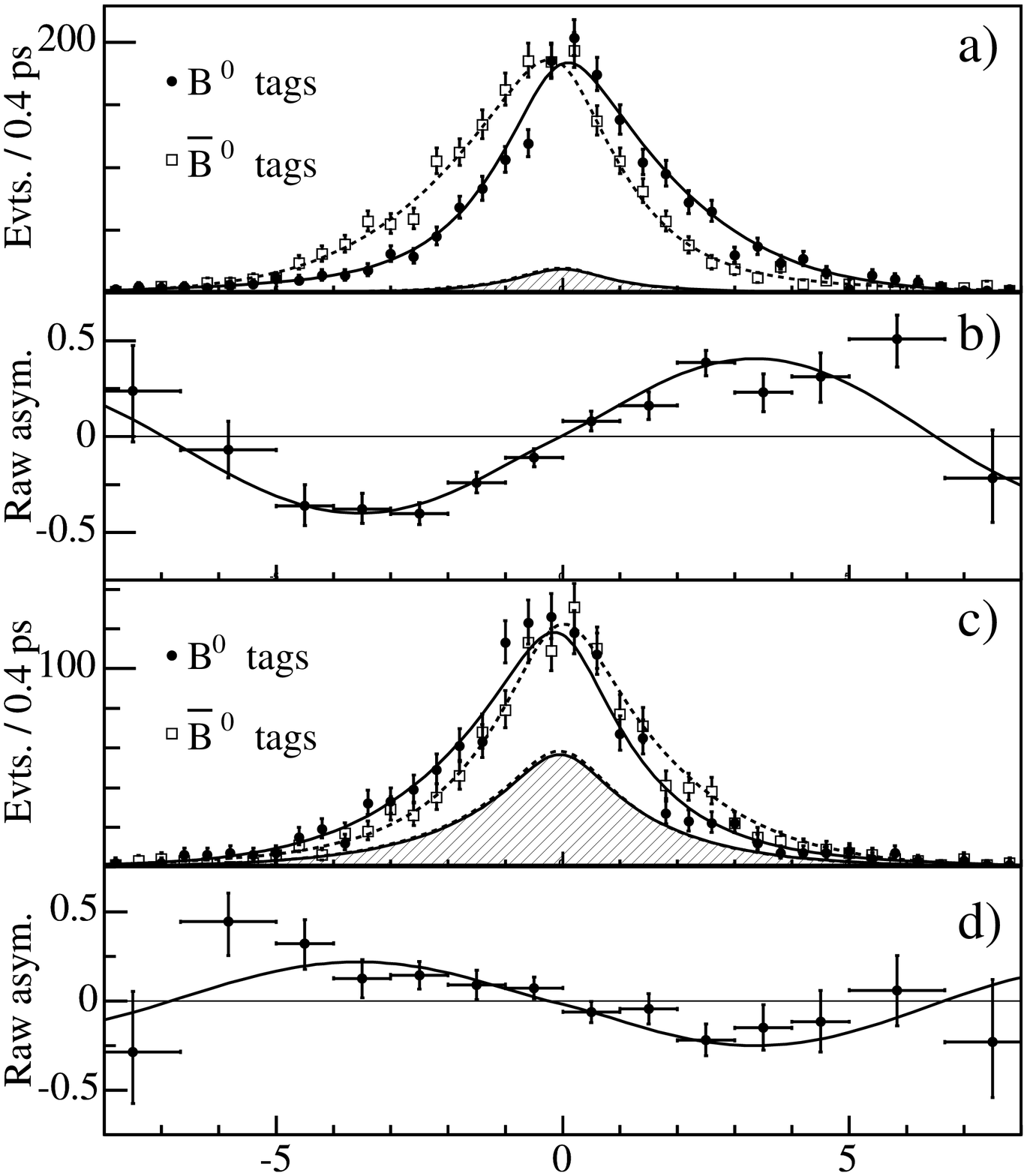,width=1.00\linewidth} 
\caption{
a) Number of $\eta_f=-1$ candidates 
($J/\psi \KS$,
$\psi(2S) \KS$,
$\chicone \KS$, and
$\eta_c \KS$)
in the signal region with a \Bz tag $N_{\Bz }$ and
with a \Bzb tag $N_{\Bzb}$, and b) the raw asymmetry
$(N_{\Bz}-N_{\Bzb})/(N_{\Bz}+N_{\Bzb})$, as functions of \deltat.
Figs. c) and d) are the corresponding plots for the $\eta_f=+1$
mode $J/\psi \KL$.  All plots exclude \othertag- tagged events.
The solid (dashed) curves represent the fit projections
in \deltat for \Bz (\Bzb) tags.
The shaded regions represent the estimated background contributions.
}
\label{fig:cpdeltat}
\end{center}
\end{figure}

The fit to the $B_{\CP}$ and \bflav\ samples yields
\begin{eqnarray}
\stwob=0.722 \pm 0.040 \stat 
\pm \syststwob \syst.\nonumber
\end{eqnarray}
\noindent
Figure~\ref{fig:cpdeltat} shows the \deltat distributions and 
asymmetries in yields between \Bz tags and \Bzb tags for the
$\eta_f=-1$ and $\eta_f = +1$ samples as a function of \deltat,
overlaid with the projection of the likelihood fit result.
\par
In a separate fit with only the high purity $\eta_f=-1$ sample,
we obtain $\vert\lambda\vert = 0.950 \pm 0.031 \stat \pm
\systal \syst$. 
The correlation between the coefficients multiplying the $\sin(\deltamd \deltat)$
and $\cos(\deltamd \deltat)$ terms in Eq.~\ref{eq:timedist} is $-2\%$.
\par
The sources of systematic error are summarized in Table~\ref{tab:systematics}.
These include
the uncertainties in the level and \CP\ asymmetry of
the peaking background,
the assumed parameterization of the \deltat\ resolution function,
possible differences between the \bflav\ and $B_{CP}$ mistag 
fractions, knowledge of the event-by-event beam spot position,
and the possible
interference between the suppressed $\bar b\to \bar u c \bar d$ amplitude with
the favored $b\to c \bar u d$ amplitude for some tag-side 
$B$ decays~\cite{dcsd}. In addition, we include the variation due
to the assumed values of 
$\vert \lambda \vert$
and $\Delta \Gamma$.
We assign the change in the measured
\stwob\ when we float $\vert \lambda \vert$
and when we set $\Delta \Gamma / \Gamma = \pm 0.02$, 
the latter being
considerably larger than recent Standard 
Model estimates~\cite{dg}. 
The total systematic error on \stwob\ ($\vert \lambda \vert$)
is $\syststwob$ ($\systal$).

\begin{table}[!htb] 
\caption{ 
Sources of systematic error on \stwob and $\vert \lambda \vert$.}
\label{tab:systematics} 
\begin{ruledtabular} 
\begin{tabular}{lcc}
 Source  &  $\sigma(\stwob)$ & $\sigma(\vert \lambda \vert)$ 
\\ \colrule
\CP\ backgrounds & 0.012 & 0.002\\
\deltat\ resolution function &0.011 & 0.003\\
$\jpsi \KL$ backgrounds &0.011 & N/A \\ 
Mistag fraction differences &0.007 & 0.001\\
Beam spot & 0.007 & 0.001\\ 
$\deltamd$, $\tau_B$, $\Delta \Gamma / \Gamma$, $\vert \lambda \vert$ &0.005 & 0.001\\
Tag-side interference & 0.003 & 0.012 \\
MC statistics & 0.003 & 0.003\\ \hline 
Total systematic error & $\syststwob$ & $\systal$\\
\end{tabular} 
\end{ruledtabular} 
\end{table}

\par
The large $B_{\CP}$ sample allows a number of consistency
checks, including separation of the data by decay mode and tagging category,
as shown in
Table~\ref{tab:result}.
Considering statistical errors only, the probability of finding 
a worse agreement in measured \stwob\ values
across decay modes is 7\% and between tagging
categories is 86\%.
The results of fits to the control samples of non-\CP decay modes
indicate no statistically significant asymmetry.
\par
This measurement of $\stwob$ supersedes our previous result~\cite{babar-stwob-prl} and is 
consistent with the range implied by other
measurements and theoretical estimates of the 
magnitudes of CKM matrix elements in the context of the Standard Model~\cite{CKMconstraints}. 
The theoretical uncertainty on the interpretation of the 
measurement of \stwob\ in these modes is 
approximately 0.01~\cite{lambda}. As the current measurement is
statistics limited, future measurements will add
further model-independent constraints on the position of the 
apex of the unitarity triangle~\cite{CKMconstraints}.  
We are grateful for the excellent luminosity and machine conditions
provided by our \pep2\ colleagues, 
and for the substantial dedicated effort from
the computing organizations that support \babar.
The collaborating institutions wish to thank 
SLAC for its support and kind hospitality. 
This work is supported by
DOE
and NSF (USA),
NSERC (Canada),
IHEP (China),
CEA and
CNRS-IN2P3
(France),
BMBF and DFG
(Germany),
INFN (Italy),
FOM (The Netherlands),
NFR (Norway),
MIST (Russia), and
PPARC (United Kingdom). 
Individuals have received support from CONACyT (Mexico), A.~P.~Sloan Foundation, 
Research Corporation,
and Alexander von Humboldt Foundation.


\begin{thebibliography}{99}

\bibitem{babar-stwob-prl}
\babar\ Collaboration, B.\ Aubert {\em et al.},
\prl {\bf 89}, 201802 (2002).

\bibitem{belle-stwob-prl}
BELLE Collaboration, K.\ Abe {\em et al.},
\pr {\bf D66}, 071102 (2002).

\bibitem{CKM}
\hyphenation{Ko-ba-ya-shi}
N.~Cabibbo, \prl {\bf 10}, 531 (1963); M.~Kobayashi and T.~Maskawa, \progtp {\bf 49}, 652 (1973).

\bibitem{BCP}
A.B.~Carter and A.I.~Sanda, \pr {\bf D23}, 1567 (1981);
I.I.~Bigi   and A.I.~Sanda, \np {\bf B193}, 85 (1981).

\bibitem{chargeconj}
Charge-conjugate reactions are included implicitly 
unless otherwise specified.

\bibitem{babar-detector-nim}
\babar\ Collaboration, B.\ Aubert {\em et al.}, 
\nim{\bf A479}, 1 (2002).

\bibitem{babar-stwob-prd}
\babar\ Collaboration, B.\ Aubert {\em et al.}, 
\pr {\bf D66}, 032003 (2002).

\bibitem{lambda}
See, for example, D.~Kirkby and Y.~Nir in S.~Eidelman {\em et al.}, \pl {\bf B592}, 1 (2004).

\bibitem{BABARTRANS}
\babar\ Collaboration, B.\ Aubert {\em et al.}. In preparation, to be submitted to \prd.

\bibitem{etacks}
\babar\ Collaboration, B.\ Aubert {\em et al.}, Submitted to \pr {\bf D},
SLAC-PUB-10368, hep-ex/0403007.

\bibitem{PDG2004}
Particle Data Group, S.~Eidelman {\em et al.}, \pl {\bf B592}, 1 (2004).

\bibitem{dcsd}
O.~Long, M.~Baak, R.~N.~Cahn, D.~Kirkby, \pr {\bf D68}, 034010 (2003).

\bibitem{dg} A.~S.~Dighe {\em et al.}, \np {\bf B624}, 377 (2002). 
M.~Ciuchini {\em et al.} JHEP {\bf 0308}, 031 (2003).

\bibitem{CKMconstraints}
See, for example, F.J.~Gilman, K.~Kleinknecht and B.~Renk in S.~Eidelman {\em et al.}, \pl {\bf B592}, 1 (2004).



\end{thebibliography}
\end{document}